\documentclass[11pt]{article}
\usepackage[utf8]{inputenc}
\usepackage[a4paper,width=165 mm, top=30mm, bottom=25mm]{geometry}
\usepackage{setspace}
\usepackage{physics}

\usepackage{amsmath}
\usepackage{amssymb}
\usepackage{authblk}

\usepackage{graphicx}
  \graphicspath{{images/}}
  \usepackage{caption}
  \usepackage{subcaption}

\usepackage{fancyhdr}
\usepackage[sorting=none]{biblatex}
\usepackage{appendix}
\usepackage{lipsum} 

\usepackage{hyperref}
    \hypersetup{
    colorlinks=true,
    linkcolor=black,
    filecolor=blue,      
    urlcolor=blue,
    citecolor=blue
    }

\title{Exploring Quantum Neural Networks for the Discovery and Implementation of Quantum Error-Correcting Codes}

\author[1]{A. Chalkiadakis\thanks{These authors contributed equally to this work}\thanks{Electronic address: \href{mailto:myronthk@gmail.com}{myronthk@gmail.com}}}
\author[1]{M. Theocharakis\protect\footnotemark[1] \thanks{Electronic address: \href{mailto:alkischalkiadakis@hotmail.com}{alkischalkiadakis@hotmail.com}}}
\author[1]{G. D. Barmparis}
\author[1,2]{G. P. Tsironis}
\affil[1]{Department of Physics, University of Crete, Heraklion 70013, Greece }
\affil[2]{John A. Paulson School of Engineering and Applied Sciences, Harvard University, Cambridge, Massachusetts 02138, USA}

\date{(Dated April 13, 2023)}
\begin{document}

\maketitle

\setstretch{1.2}

\section{Abstract}
We investigate the use of Quantum Neural Networks for discovering and implementing quantum error-correcting codes. Our research showcases the efficacy of Quantum Neural Networks through the successful implementation of the Bit-Flip quantum error-correcting code using a Quantum Autoencoder, effectively correcting bit-flip errors in arbitrary logical qubit states. Additionally, we employ Quantum Neural Networks to restore states impacted by Amplitude Damping by utilizing an approximative 4-qubit error-correcting codeword. Our models required modification to the initially proposed Quantum Neural Network structure to avoid barren plateaus of the cost function and improve training time. Moreover, we propose a strategy that leverages Quantum Neural Networks to discover new encryption protocols tailored for specific quantum channels. This is exemplified by learning to generate logical qubits explicitly for the bit-flip channel. Our modified Quantum Neural Networks consistently outperformed the standard implementations across all tasks.

\section{Introduction}

Machine learning (ML) and its extension, deep learning, are subfields of artificial intelligence that enable knowledge acquisition through experience rather than explicit instructions. Neural networks, the foundation of deep learning, have become integral to our daily lives. As AI methods are increasingly employed to address complex problems involving large volumes of data, researchers are exploring the potential of neural networks in tackling contemporary scientific challenges.

Quantum computing has recently emerged as a highly promising research area, sparking efforts to integrate this potentially transformative technology with artificial intelligence algorithms \cite{QML}, including neural networks. Classical neural networks (cNNs) have demonstrated remarkable capabilities in machine learning, with quantum counterparts holding the promise of handling complex tasks involving unknown quantum algorithms. By leveraging the back-propagation algorithm, neural networks can identify correlations among intricate data points and extract valuable information, often yielding results unattainable through other means. This foundation has led to the development of QNNs \cite{speedup, hybrid_qnn,q_supp_vec_mac}, which aim to match or surpass cNNs by utilizing the theoretically superior power of quantum computing devices \cite{arute2019quantum}. In this paper, we will explore an implementation introduced in \cite{beer2020training} where the authors demonstrated a method for efficient training of so-called
dissipative quantum neural networks (DQNNs) on training data pairs in
form of input and desired output quantum states. This version of QNNs acts as direct analogs of fully-connected feed-forward NNs, which  trace out qubits from previous layers during the transition to new layers, resulting in energy dissipation, which gave it their name.

Prior work on quantum denoising with quantum autoencoders (QAEs) \cite{q_auto_mat, achache2020denoising} has showcased their ability to denoise specific quantum states, like Greenberger–Horne–Zeilinger (GHZ) states, affected by specific quantum noise. The authors have demonstrated the ability of QAEs in reconstructing noisy states as well as generating noise-free states. Expanding on this work, our objective is to successfully implement general error-correcting codes using QAEs in arbitrary logical qubit states.

In the following work, we successfully implement the Bit-Flip quantum error-correcting code using a Quantum Autoencoder, demonstrating its ability to correct bit-flip errors in arbitrary logical qubit states. Additionally, we apply QNNs to error-correct states affected by Amplitude Damping, utilizing an approximate 4-qubit error-correcting codeword \cite{ampl_dump_1} \cite{ampl_dump_2}. Lastly, we propose a strategy employing QNNs to discover new encryption protocols for specific quantum channels. We use a Quantum Neural Network to learn the creation of logical qubits tailored for the bit-flip channel. However, we find that our models can only accomplish this task after modifying their structure. These modifications accelerate the training procedure and help the models avoid barren plateaus of the cost function observed in previous work, ultimately improving training task performance. Our results indicate that modified DQNNs outperform default implementations across all tasks considered in this work.

\section{Quantum Neural Network’s Architecture}

In this section, we discuss the architecture of QNNs and their implementation. By representing a fully connected feed-forward QNN with a series of consecutive quantum operations, we describe the action of each layer and how the quantum states of the qubits in each layer are obtained. Additionally, we outline the training process for the network, including the evaluation of its performance through a cost function that measures the distance between input and output states. 

We simulated the quantum circuits that run those networks using python's package qiskit, assuming that the qubits are noisy-free in general. The code to do so was provided by the authors in \cite{achache2020denoising} and has been upgraded for the purposes of this paper, and the nets now include more optimisers\footnote{Namely, we added RMSprop, Adamax and Nadam into the already existing ones which were SGD, Adam.}, early stopping, gradient ascent as well as the ability to implement Conjugate Layers (introduced in later section).

\begin{figure}[h!]
    \centering
    \includegraphics[width=0.8\textwidth]{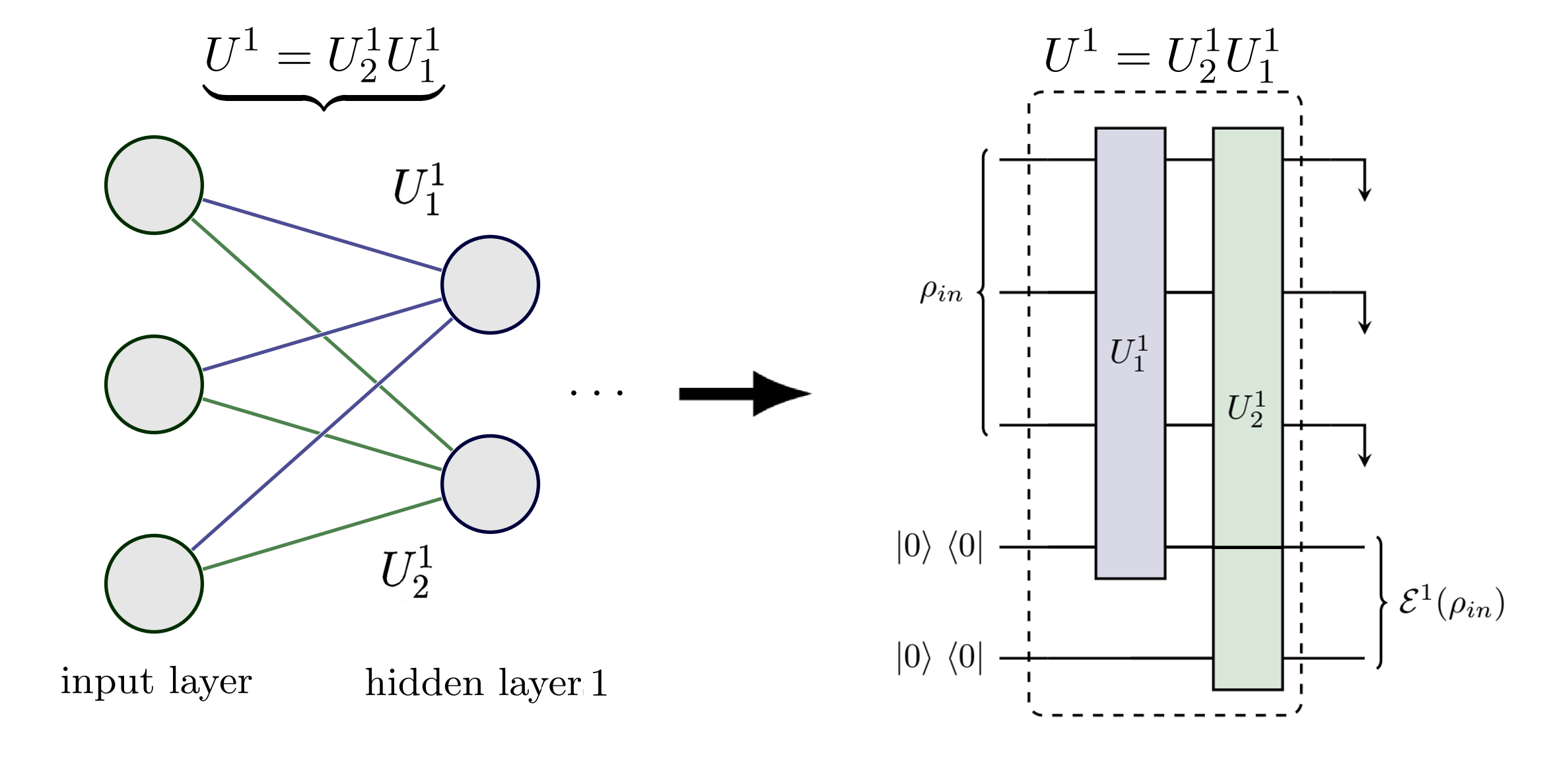}
    \caption{(Left) Schematic representation of the first layer of a $\qty[3, 2,...]$ QNN. (Right) Quantum circuit that constructs the first layer of that QNN. Note that the unitary $U^1_2$ acts on the first 3 and last qubits only. }
\end{figure}

We denote by $[m_{in},m_1, . . . , m_L,m_{out}]$  a fully connected feed-forward QNN with $L$ hidden layers, with each having $m_l$ number of neurons. Then, a QNN simply represents a series of consecutively quantum operations, with its output state being 
\begin{equation}
\rho^{\text {out }}=\mathcal{E}\left(\rho^{\text {in }}\right)=\mathcal{E}^{out}\left(\mathcal{E}^L\left(\ldots \mathcal{E}^2\left(\mathcal{E}^1\left(\rho^{\text {in }}\right)\right) \ldots\right)\right)
\end{equation}
where the map of layer $l$, $\mathcal{E}^{l}$, is defined via 
\begin{equation}
\mathcal{E}^l\left(\rho ^{l-1}\right) \equiv \operatorname{Tr}_{l-1}\left(U^l\left(\rho^{l-1} \otimes|0 \ldots 0\rangle_l\langle 0 \ldots 0|\right) (U^{l}) ^ \dagger\right)
\end{equation}
with $U^l \equiv \prod_{j=1}^{m_l} U_j^l$. The  unitary operator, $U_j^l$, connects all the qubits of the $l - 1$ layer with the $j^{th}$ qubit of layer $l$. The quantum state of the qubits in layer $l$ is therefore obtained by applying, in ascending order of j, all the unitaries $U^l_j$ and then tracing out all qubits of the previous layer.

The authors of \cite{QNN} have demonstrated that this structure is capable of simulating a universal quantum computer \cite{deutsch1985quantum}, which effectively means that any quantum algorithm can be built with this method given enough resources. This is certainly one of the most important features of this implementation of QNNs.

To train the network, in a supervised manner, one has to have access to N input/target training pairs of the form $\qty{ \ket{\psi^{in}_x}, \ket{\psi^{targ}_x}} _{x \in {1,2,...,N}}$ that for concreteness we can assume to take the form, $\ket{\psi^{targ}_x} = V \ket{\psi^{in}_x}$, where V is an unknown unitary operation that the QNN has to replicate. 

Then, to evaluate its performance we define a cost (loss) function that returns the distance between the input and the output states. One natural choice is the Fidelity, $F\qty(\ket{\psi^{targ}_x}, \rho^{out}_x )$, averaged over all the training pairs
\begin{equation}
    C(\boldsymbol{\kappa}) = \frac{1}{N}\sum_{x=1}^N \bra{\psi^{targ}_x} \rho^{out}_x \ket{\psi^{targ}_x}
\label{cost_fucntio}
\end{equation}
where $\boldsymbol{\kappa}$ is a vector that contains all the parameters of the network. The cost function takes a value of 1 when the target and output states are all the same and 0 when they are all perpendicular to each other. Naturally, to train the network we have to maximize the cost function by applying a gradient ascent (instead of decent) algorithm\footnote{Sometimes as a cost function is considered to be $1-C$ so that we can use gradient descent algorithms that are more familiar} 
\begin{equation}
\boldsymbol{\kappa}^{t+1} = \boldsymbol{\kappa}^{t}+\eta \grad_{\boldsymbol{\kappa}} C(\boldsymbol{\kappa}^t)
\end{equation}
where $t$ denotes the training step or the epoch and $\eta$ the learning rate, typically a small number that ensures that the gradient step is in the vicinity where the cost function decreases

Moreover, the unitary transformations are parameterized in the following way 
\begin{equation}
U_j^l=e^{i K^l_j}
\end{equation}
with 
\begin{equation}
K^l_j=\sum_{\sigma \in P ^{ \otimes(m_l+1)}} k_{\sigma} \cdot \sigma
\label{kappa}
\end{equation}
where $k_\sigma$ are real numbers and the parameters to be learned and $P^{\otimes j}$ is the set of all possible tensor products of
length $j$ between the elements of $P=\{I, X, Y, Z\}$, i.e. $P^{\otimes 2}=\{I I, I X, I Y, I Z, X I, X X \ldots\}$. In this way, $U_j^l$ are uniquely defined
by their coefficients $k_\sigma$. 

Between two layers with $m_l$ and $m_{l+1}$ number of qubits each, we have $m_{l+1}$ unitary transformations that have $4^{m_l +1}$ number of coefficients. So in total, the number of trainable coefficients that a QNN has, is
\begin{equation}
\sum_{i=1}^{\ell-1} m_{i+1} \cdot 4^{m_i+1}
\label{num_of_params}
\end{equation}
As one can observe, this number scales exponentially fast with the number of qubits in a layer. For this reason, it becomes unpractical to simulate QNNs with classical computers, which have more than a handful of qubits. Furthermore, at any point in the QNN, to construct the quantum map $\mathcal{E}^l$, we reuse qubits from previous layers by resetting them to the state $\ket{0}$, in order to use the least amount of qubits possible. The estimation of the cost function (\ref{cost_fucntio}) is done by implementing the swap test algorithm \cite{barenco1997stabilization} between the output state of the QNN and
the target state, which deduces their closeness, (see Appendix A for more information).

In short, by employing a series of consecutive parameterized gates between layers and resetting neurons to $\ket{0}$ state when transitioning to a new layer we can construct a Quantum Neural Network. Through a supervised training method that utilizes gradient ascent to maximize the cost function, which is chosen to be the Fidelity between the output of the QNN and a target state of our choice, QNNs can effectively learn to perform unknown quantum algorithms.

\subsection{Enhancing QNN Performance with Conjugate Layers}
We'll now introduce the concept of conjugate layers in QNNs, which modify the original architecture of those models to address the challenges associated with training plateaus and training time.

Restrictions imposed during the training process in classical Neural Networks can often lead to improved performance and reduced training time. For example, techniques such as dropout \cite{srivastava2014dropout} and U-nets \cite{ronneberger2015u} resolve issues like overfitting and enable the transfer of information between different parts of the network. In their quantum version, conjugate layers serve as a quantum analog of these restrictions, offering similar benefits.

The concept behind conjugate layers is straightforward: a conjugate layer can be implemented by replacing the transformation of a layer, $l$, with the Hermitian conjugate of the unitary transformation of a different layer of choice, call it $l'$
\begin{equation}
    U^l_{conj} = (U^{l'})^{\dagger}= \qty(\prod_{j=1}^{m_{l'}} U_j^{l'})^{\dagger} = \prod_{j=m_{l'}}^{1} (U_j^{l'})^{\dagger}
\end{equation}
where 
\begin{equation}
    (U_j^{l'})^{\dagger}=e^{-i K^{l'}_j}
\end{equation}
Thus, these two layers are trained simultaneously; and in doing so we essentially truncate the number of trainable parameters of the QNN as the parameters comprising layer $l$ now also characterize layer $l'$.

To implement this method effectively, the conjugate layer must apply the conjugate unitary transformations in reverse order. For instance, if layer $l$ contains $m$ neurons that are transformed into $n$ in the subsequent layer, the conjugate layer should have $n$ neurons that are transformed back into $m$. As a result, the QNN will exhibit the following structure:
$$[...,\underbrace{m,n}_{U^l},..., \underbrace{n,m}_{U^l_{conj} },...]$$
This approach enables the integration of multiple conjugate layers within the QNN architecture, accommodating up to $q$ conjugate layers in a QNN with a total of $2q+1$ layers.

\begin{figure}[h!]
    \centering
    \includegraphics[width = 0.49\textwidth]{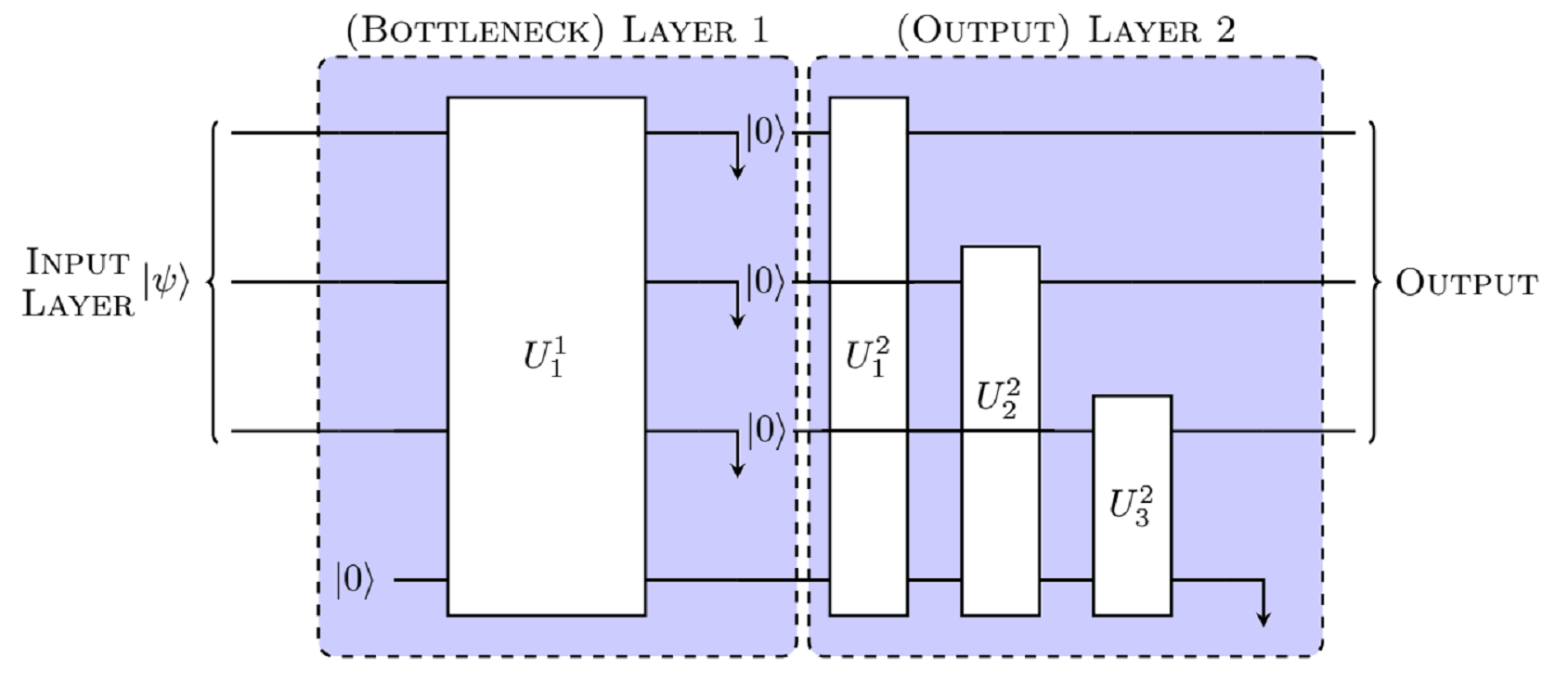}
    \includegraphics[width=0.49\textwidth]{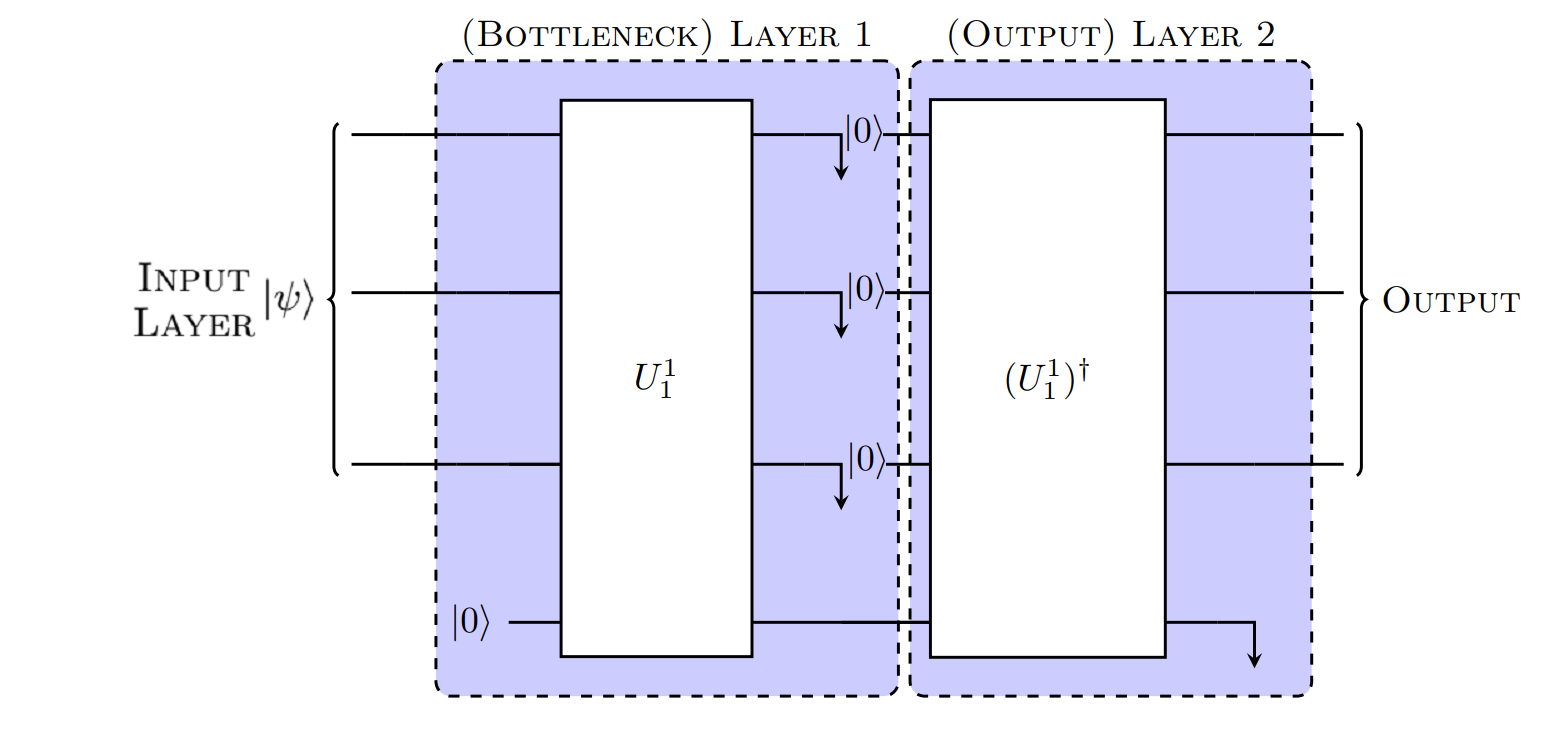}
    \caption{(Left) Vanilla [3,1,3] QNN. (Right) [3,1,3] QNN with a conjugate layer. }
    \label{fig313}
\end{figure}

In this example, we showcase two versions of the same QNN employed for error-correcting Bit Flip errors. In Figure (\ref{fig313}), the left circuit represents the vanilla QNN implementation, while the right one replaces the transformation of the output layer with the Hermitian conjugate of the first layer. Observe how the three two-qubit unitary transformations, $U^2_{1,2,3}$, are replaced by a single 4-qubit transformation $(U^1_1)^\dagger$.

\section{Quantum Autoencoders for Bit Flip Error-Correction}
Classical autoencoders are typically employed for removing unwanted features, e.g. noise, from the dataset. Quantum Auto-Encoders (QAE) are very much inspired by their classical counterpart and can be set to perform quantum error correction by removing the least relevant features from quantum states, like quantum noise. Unlike conventional quantum error correction algorithms that typically require generalized measurements and classical information processing, QAEs are supposed to be capable of performing those tasks autonomously. 

In this section, we will investigate if QAEs are able to denoise specific types of errors,  by utilizing Quantum Error Correcting Codes. The general approach, is to fit the QAE multiple different encoded states, which are corrupted by a quantum channel, denoted as $\tilde{\ket{\psi}}_L$, and then train the QAE to replicate the states with no error, $\ket{\psi}_L$.

Our initial focus is on correcting errors generated by the Bit-flip channel. The bit flip channel applies a NOT gate to every qubit with probability $p$, and is described by the following operation elements:
\begin{equation}
E_0=\sqrt{1-p} I=\sqrt{1-p}\left[\begin{array}{ll}
1 & 0 \\
0 & 1
\end{array}\right] \quad E_1=\sqrt{p} X=\sqrt{p}\left[\begin{array}{ll}
0 & 1 \\
1 & 0
\end{array}\right]
\end{equation}
$I$ is the identity matrix which corresponds to the case where qubit state was left uncorrupted and X is the
Pauli matrix $\sigma_x$ responsible for the bit flip. To do so, we utilized the 3-qubit error-correcting code by encoding all of the input/target states as follows
\begin{equation}
    \ket{\psi} = a\ket{0}+b\ket{1} \rightarrow \ket{\psi}_L = a\ket{000}+b\ket{111}
\end{equation}

In order to train QAEs to denoise bit flips, we created a training set of 120 input/target pairs of the form $\qty{\tilde{\ket{\psi}}_L,\ket{\psi}_L}$, where $\ket{\psi}_L$ is one of the following states: $\ket{0}_L$, $\ket{1}_L$, $\ket{+}_L$, $\ket{-}_L$, $\ket{+i}_L$, $\ket{-i}_L$\footnote{$\ket{+}=\frac{\ket{0}+\ket{1}}{\sqrt{2}}$ and $\ket{-} = \frac{\ket{0}-\ket{1}}{\sqrt{2}}$.}. The input states of the set are corrupted by (single qubit) bit-flip errors, with probability $p=0.2$. Additionally, the models were trained in training sessions. Each session had fixed hyperparameters and continued the training from the model of the previous session. Initial sessions generally had a relatively large learning rate ($lr= 0.25$) with Adam or Nadam optimizer and a batch size of 20. Latter sessions had decreased lr, vanilla SGD optimizer, and no batch size. This strategy was adapted such that we ensure convergence of the cost function as well as to save computational time. Lastly, the initialization of the QNN's parameters was chosen to be uniformly random between 0 and 1, rather than all being fixed to 0 (as done in \cite{qnn-concept}), to avoid a plateaued starting point.
\begin{figure}[h!]
    \centering
    \includegraphics[width = 0.6\textwidth]{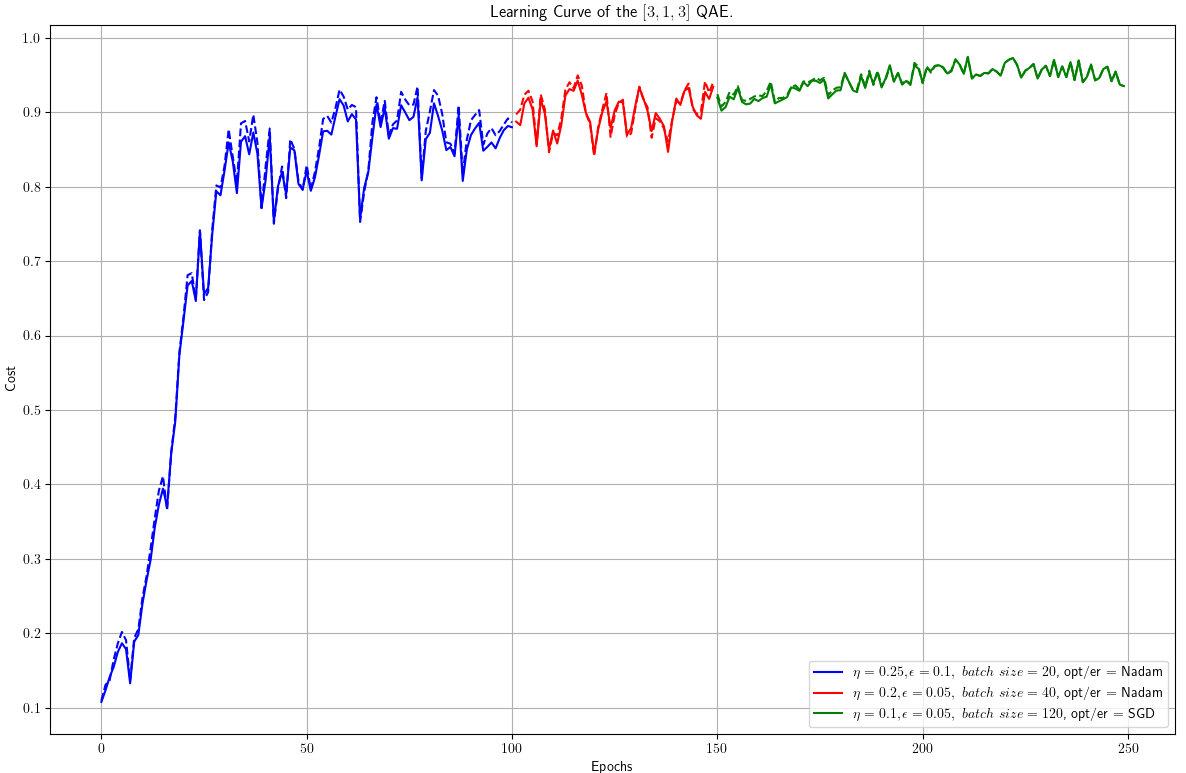}
    \caption{Learning Curve of the $[3,1,3]$ QAE. The dashed line indicates the cost function on the training set, while the continued one, on the validation set. Each color represents a learning session with fixed hyperparameters. The training continues for another 250 epochs (500 in total) but the model showed minimal improvement in those last epochs.}
    \label{learning_313}
\end{figure}

To evaluate the model's performance we created a validation set during training, which ensures that the QAE generalizes and corrects bit-flip errors from any arbitrary state. To this end, we parameterized the (logical) qubit's state in the Bloch sphere representation 
\begin{equation}\label{Miri31}
    \ket{\psi(\theta,\phi)}_{L} = cos(\theta/2) \ket{0}_L + e^{i\phi} sin(\theta/2) \ket{1}_L.
\end{equation}
Then created a meshgrid with $N=20$ different values for the parameters $\theta$ and $\phi$, where $\theta \in [0,\pi]$ and $\phi\in[0,2\pi]$ . The validation set, thus, contained in total $400(=20\times20)$ different states, such that they uniformly cover the whole Bloch sphere. Those states were then corrupted with a bit-flip error with $p=0.2$. The mean Fidelity($\equiv$ Cost of the validation set) is calculated by 
\begin{align}
    \Bar{F} &= \frac{1}{4\pi}\int_0^{2\pi} \int_0^{\pi} F(\rho_{out}(\theta,\phi),\ket{\psi(\theta,\phi) }_{targ}) sin(\theta) d\theta d\phi \\ 
    &\simeq \frac{1}{4\pi} \sum_{i=0}^{N} \sum_{j=0}^{N} F(\rho_{out}(\theta_i,\phi_j),\ket{\psi(\theta_i,\phi_j)}_{targ} ) sin(\theta_i) \Delta \theta \Delta \phi \label{eq11}
\end{align}
with $\Delta \theta = \pi/ N$,  $\Delta \phi = 2\pi/ N$, $\theta_i = i \Delta \theta$ and $\phi_j = j \Delta \phi$.

Following this strategy, we began training multiple models with different hyperparameters each. Figure (\ref{learning_313}) depicts the learning period of the first 250, out of 500, epochs of the best-performing model. As we can observe overfitting isn't a concern here; the QAE successfully learns to denoise bit-flips, with the Cost function on both the train and the validation set eventually reaching a value near 1, at around 0.98.

Continuing with the testing of the model, it would be useful to study the performance of the model given that a bit flip took place in one of the 3 qubits that comprise the logical qubit or that no error has occurred. To achieve this, we calculated the corresponding conditional Fidelities, on 4 validation sets
\begin{align*}
    \Bar{F}(\rho_{out},\ket{\psi}| \text{bit-flip to qubit 1} ) = 0.96 \\ 
    \Bar{F}(\rho_{out},\ket{\psi}|\text{bit-flip to qubit 2} ) = 0.96 \\
    \Bar{F}(\rho_{out},\ket{\psi}|\text{bit-flip to qubit 3} ) = 0.95 \\
    \Bar{F}(\rho_{out},\ket{\psi}|\text{no error} ) = 0.97
\end{align*}
This confirms that the QAE has successfully learned to denoise single bit-flip events on  (logical) qubits, with high fidelity. 

\begin{figure}[h!]
    \centering
    \includegraphics[width=0.45 \textwidth]{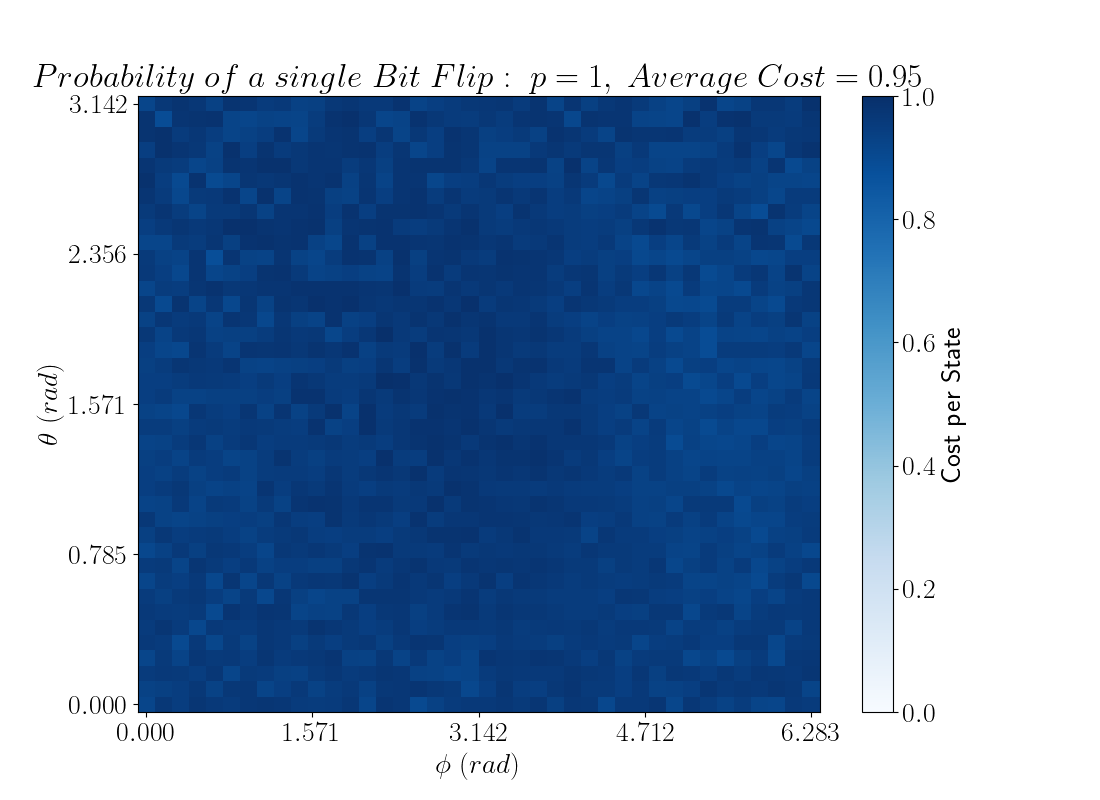}
    \includegraphics[width=0.45 \textwidth]{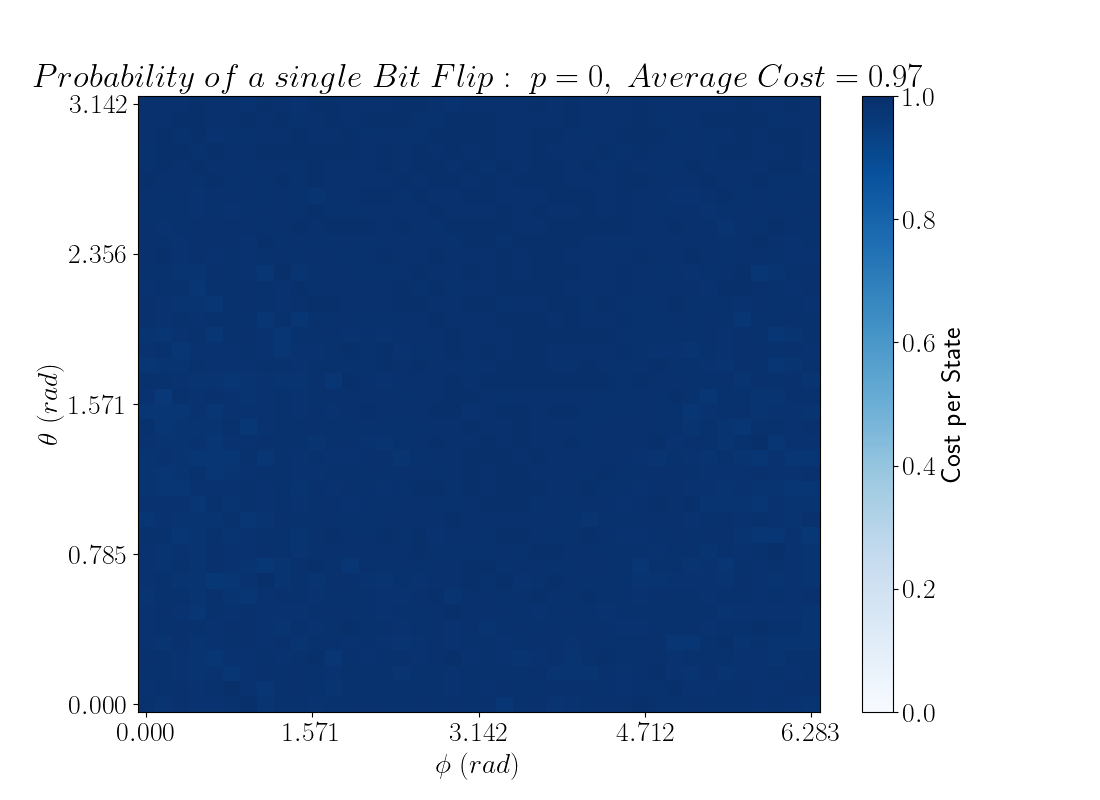}
    \caption{Each point in the graphs corresponds to a qubit state on the Bloch sphere (\ref{Miri31}). The Autoencoder was tested with $N=1600$ different qubit states.}
    \label{fig4}
\end{figure}

To this end, we present the colormaps in figure (\ref{fig4}) that depict the fidelity $F(\rho(\theta,\phi), \ket{\psi(\theta,\phi)})$ as a function of the parameters $\theta$ and $\phi$ on two different validation sets. In the first validation set, the probability of error was set to be $p=1$, while in the second one $p=0$.

\begin{figure}[h!]
    \centering
    \includegraphics[width = 0.4\textwidth]{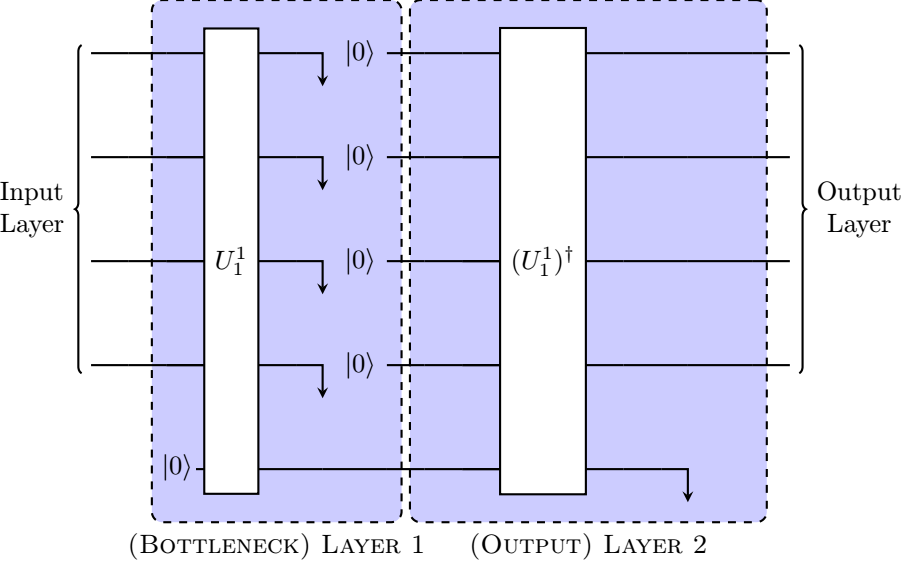}
    \caption{Architecture of Autoencoders used for error-correcting the Amplitude Damping.}
    \label{fig5}
\end{figure}

With this method, we essentially check if the QAE has developed  a bias during training towards any specific region on the Bloch sphere. The colormaps would reveal this kind of bias if their color were not uniform. In such cases, the training set must be accordingly modified and include a representative state from that region so that the Autoencoder can generalize more efficiently. Figure (\ref{fig5}) suggests that there are no areas on the Bloch sphere where our model underperforms.   

\subsubsection*{Limitations of this Approach}

While employing pre-existing error-correcting codes in QAE models can be advantageous, it is important to recognize that these models also inherit the limitations of such methods. For instance, if more than two bit flip errors occur, the QAE may fail to accurately identify and correct the error, leading to improper error correction for that particular state. In an optimal quantum error-correcting algorithm, the probability of successful correction is given by:

\begin{equation}
3(1-p)p^2+p^3
\end{equation}

Although the QAE models developed in this work are sub-optimal algorithms, they still demonstrate the ability to recover corrupted states with high fidelity. As a result, their probability of successful correction closely follows the same dependence as optimal quantum error-correcting algorithms. However, it is essential to be aware of these limitations when employing QAEs for error correction tasks, as they may impact the overall performance of the system.

\section{Addressing Amplitude Damping Channel Errors with QAEs}

While the Bit flip channel serves as an instructive starting point for understanding and testing new error-correcting techniques, its simplicity may not reveal the full extent of potential limitations inherent to the chosen method. Consequently, we have chosen to explore the use of QAEs to correct errors in the Amplitude Damping channel, which describes spontaneous emission in quantum systems and represents a more realistic type of quantum noise. Additionally, efficiently correcting Amplitude Damping errors requires a 4-qubit error-correcting code, further challenging our models.

For a single qubit state the Kraus operators that describe Amplitude Damping are
\begin{equation} \label{eq6.14}
    E_{0}=\left[\begin{array}{cc}1 & 0 \\ 0 & \sqrt{1-\gamma}\end{array}\right], \quad
    E_{1}=\left[\begin{array}{cc}0 & \sqrt{\gamma} \\ 0 & 0\end{array}\right]
\end{equation}
where $\gamma$ is the probability of our state emitting a photon. This probability is considered to be small in order to ensure weak interactions of our qubit with its environment. The operator $E_1$ describes the "damping event", where the state $\ket{1}$ decays to $\ket{0}$ by emitting a photon. The operator $E_0$ leaves $\ket{0}$ unchanged, but reduces the amplitude of $\ket{1}$. Physically this implies that there was no "damping event", thus the environment perceives that the state of our qubit is more likely to be $\ket{0}$, rather than $\ket{1}$.

In contrast to the bit flip channel, finding an appropriate error-correcting code for efficiently addressing Amplitude Damping is considerably more challenging, as none of the Kraus operators leave the qubit state uncorrupted. Consequently, each real qubit within the error-correcting code will be corrupted by either operator $E_0$ or $E_1$, making it impossible to identify and correct all possible corruptions of Amplitude Damping using simple error-correcting codes.

However, by keeping $\gamma$ small, the perturbation caused by $E_0$ on our qubit state becomes negligible. In this scenario, we can design approximate error-correcting codes that disregard the action of $E_0$ on our qubit and concentrate on correcting corruptions resulting from the action of $E_1$ that describes the spontaneous decay of a quanta of energy. One such error-correcting code, proposed in \cite{ampl_dump_2}, employs four real qubits to correct single qubit "damping events".

This 4-qubit error-correcting code suggests that an uncorrupted single qubit state should be encoded as
\begin{equation}\label{eq17}
    \ket{\psi} = a \ket{0} + b \ket{1} \rightarrow \ket{\psi}_L = \frac{a}{\sqrt{2}}\left(|0000\rangle+|1111\rangle \right) + \frac{b}{\sqrt{2}}\left(|0011\rangle+|1100\rangle \right).
\end{equation}

In the case where we only allow single qubit "damping events", and we use $\ket{\psi}_L$ as an input state for the Amplitude Damping channel, the output of this quantum operation can be one of the following states
\begin{equation} \label{eq6.19}
\begin{aligned}
&\left|\phi_{0000}\right\rangle=a\left[\frac{|0000\rangle+(1-\gamma)^2|1111\rangle}{\sqrt{2}}\right] + b\left[\frac{(1-\gamma)[|0011\rangle+|1100\rangle]}{\sqrt{2}}\right], \\
&\left|\phi_{1000}\right\rangle=\sqrt{\frac{\gamma(1-\gamma)}{2}}[a(1-\gamma)|0111\rangle+b|0100\rangle], \\
&\left|\phi_{0100}\right\rangle=\sqrt{\frac{\gamma(1-\gamma)}{2}}[a(1-\gamma)|1011\rangle+b|1000\rangle], \\
&\left|\phi_{0010}\right\rangle=\sqrt{\frac{\gamma(1-\gamma)}{2}}[a(1-\gamma)|1101\rangle+b|0001\rangle], \\
&\left|\phi_{0001}\right\rangle=\sqrt{\frac{\gamma(1-\gamma)}{2}}[a(1-\gamma)|1110\rangle+b|0010\rangle] .
\end{aligned}
\end{equation}
where the squares of the norm of these states give their probabilities for occurring in a mixture. The subscript signifies which operator, $E_0$ or $E_1$, has been applied to the corresponding real qubit of the state presented in equation (\ref{eq17}).

To error-correct Amplitude Damping, we trained QAEs following a strategy similar to that used for the bit flip channel. However, due to some crucial differences between these two quantum channels, certain adaptations to the training procedure were necessary.

First and foremost, for our approximate error-correcting code to be valid, the probability $\gamma$ of the "damping event" must be kept relatively small. As a consequence, the majority of the corrupted states encountered by the Autoencoder during training will be described by the general quantum state $\ket{\phi_{0000}}$, where no "damping event" has occurred. This leads the Autoencoder to develop an inevitable bias towards that quantum state. To mitigate this issue, we introduced a second probability $p$ that allows us to control the frequency with which each corruption is included in the training set, regardless of the value of $\gamma$. Lower values of $p$ result in training sets that predominantly contain states described by $\ket{\phi_{0000}}$, whereas higher values of $p$ yield training sets that mostly consist of states of the form $\ket{\phi_{1000}}$, $\ket{\phi_{0100}}$, $\ket{\phi_{0010}}$, and $\ket{\phi_{0001}}$.

The training process for the models was conducted in multiple sessions. During each session, we employed the default optimizer, no batch size, and maintained the probability $\gamma$ at $0.1$. Initial training sessions utilized training sets containing 50 input/target state pairs in the form of ${ \tilde{\ket{\psi}}_L, \ket{\psi}_L }$, where $\ket{\psi}_L$ could be one of the following states: $\ket{0}_L$, $\ket{1}_L$, or $\ket{+}_L$. Additionally, we set the probability p to $0.2$ and the learning rate to $0.2$. In later stages of training, we increased the total number of state pairs to 70\footnote{Due to computational power constraints, we had to limit ourselves to training sets comprised of no more than 70 state pairs.} and replaced some initial training set states with representative states from regions of the Bloch sphere where our models faced difficulties. Furthermore, we increased the probability p to $0.8$ and decreased the learning rate to $0.1$.

The validation set that we used in order to supervise the performance of our models was comprised by 400 different states in total. These states were parameterized in the Bloch sphere representation (equation \ref{Miri31}) just like the case of the Bit Flip channel and we chose 20 values for the parameter $\theta \in [0, \pi]$ and 20 values for $\phi \in [0,2\pi]$.

The architecture of the Quantum Autoencoders (QAEs) we trained also exhibited some differences. In this case, we were required to use 4 qubits in the input layer, as our encoded state consists of 4 qubits, 1 qubit in the latent space, and 4 qubits in the output layer, resulting in a [4,1,4] Autoencoder. These models also incorporated conjugate layers. Specifically, the unitary operator of the output layer was configured as the Hermitian conjugate of the input layer's unitary operator.

\begin{figure}[h!]
    \centering
    \includegraphics[width = 0.49\textwidth]{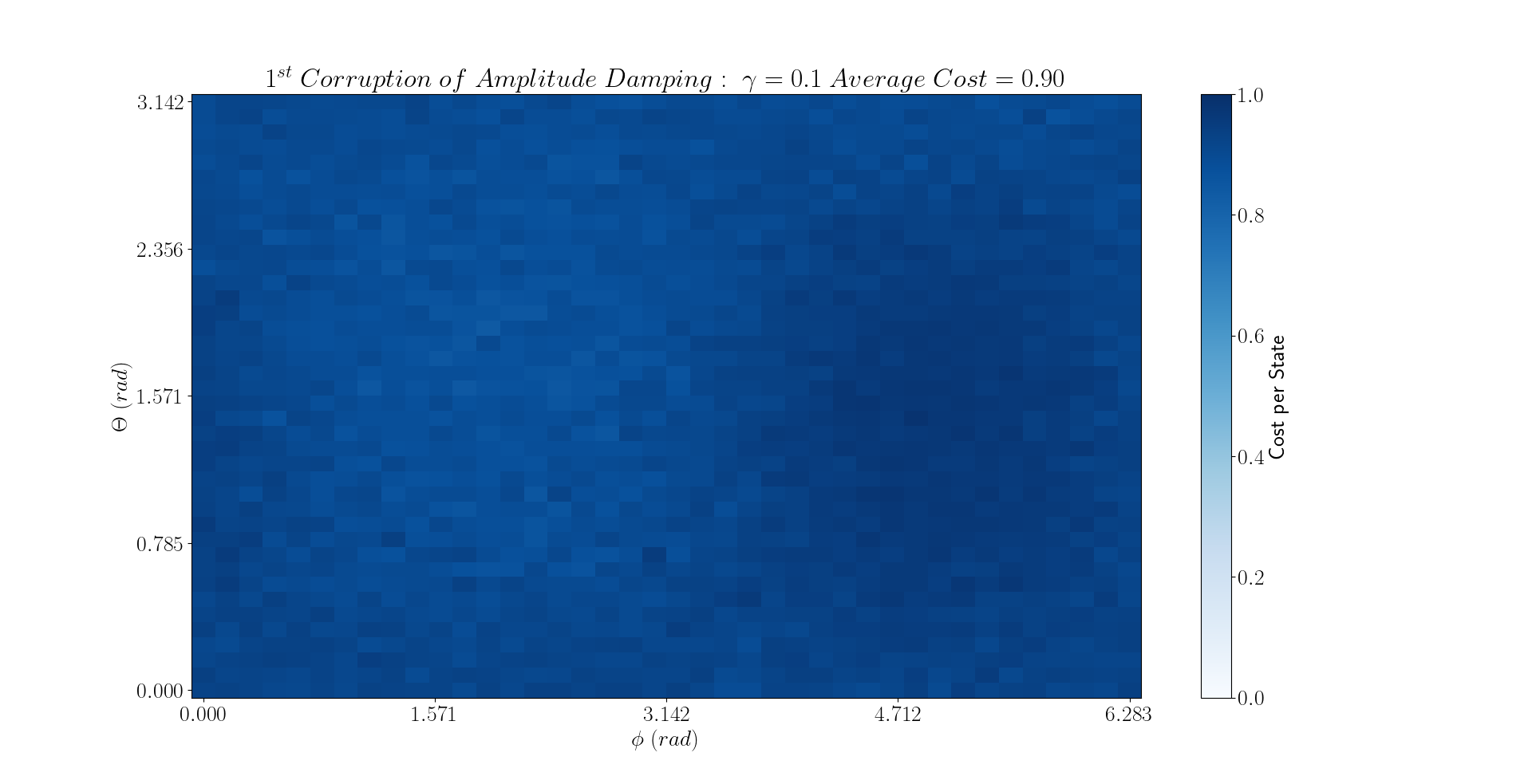}\\
    \includegraphics[width = 0.49\textwidth]{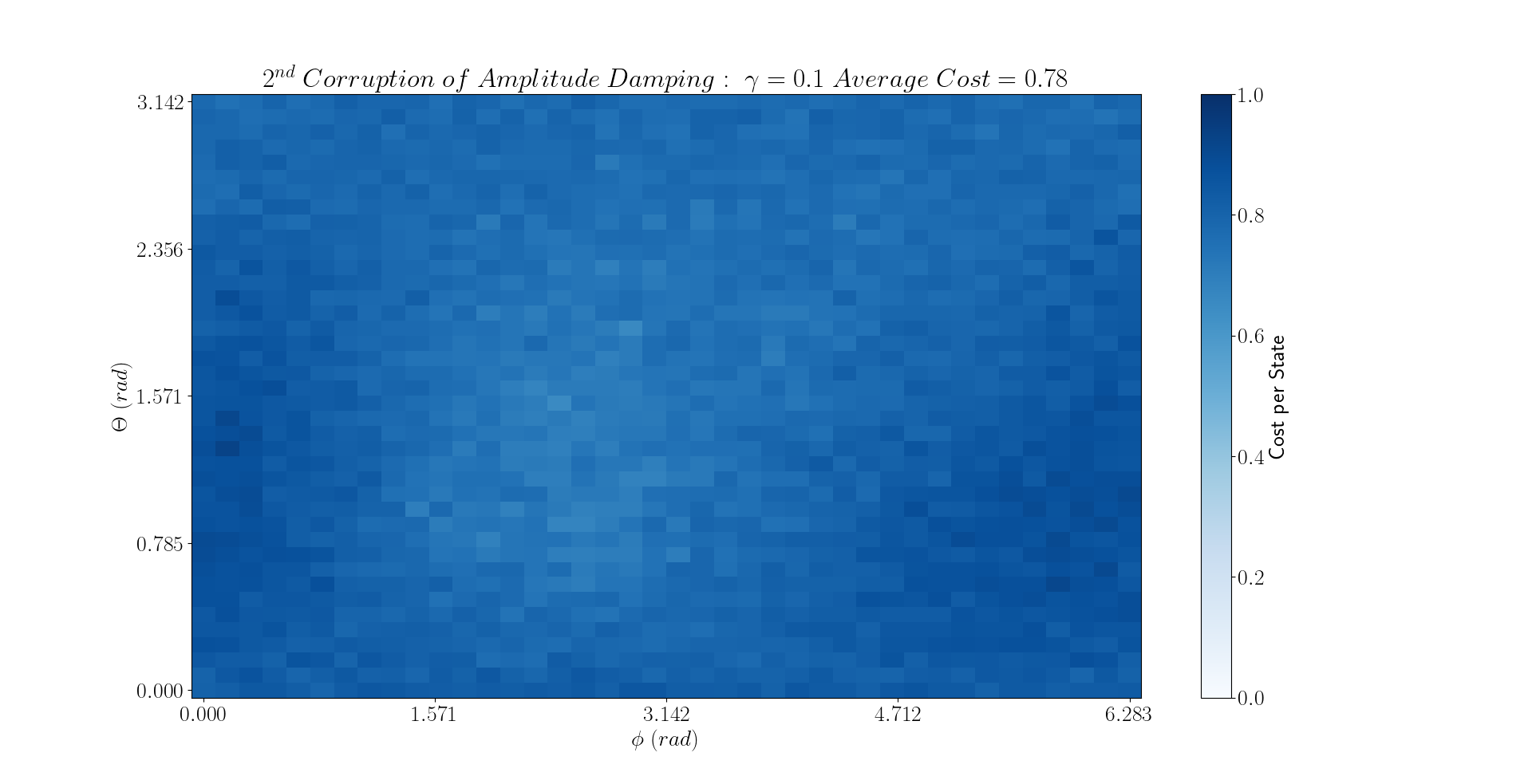}
    \includegraphics[width = 0.49\textwidth]{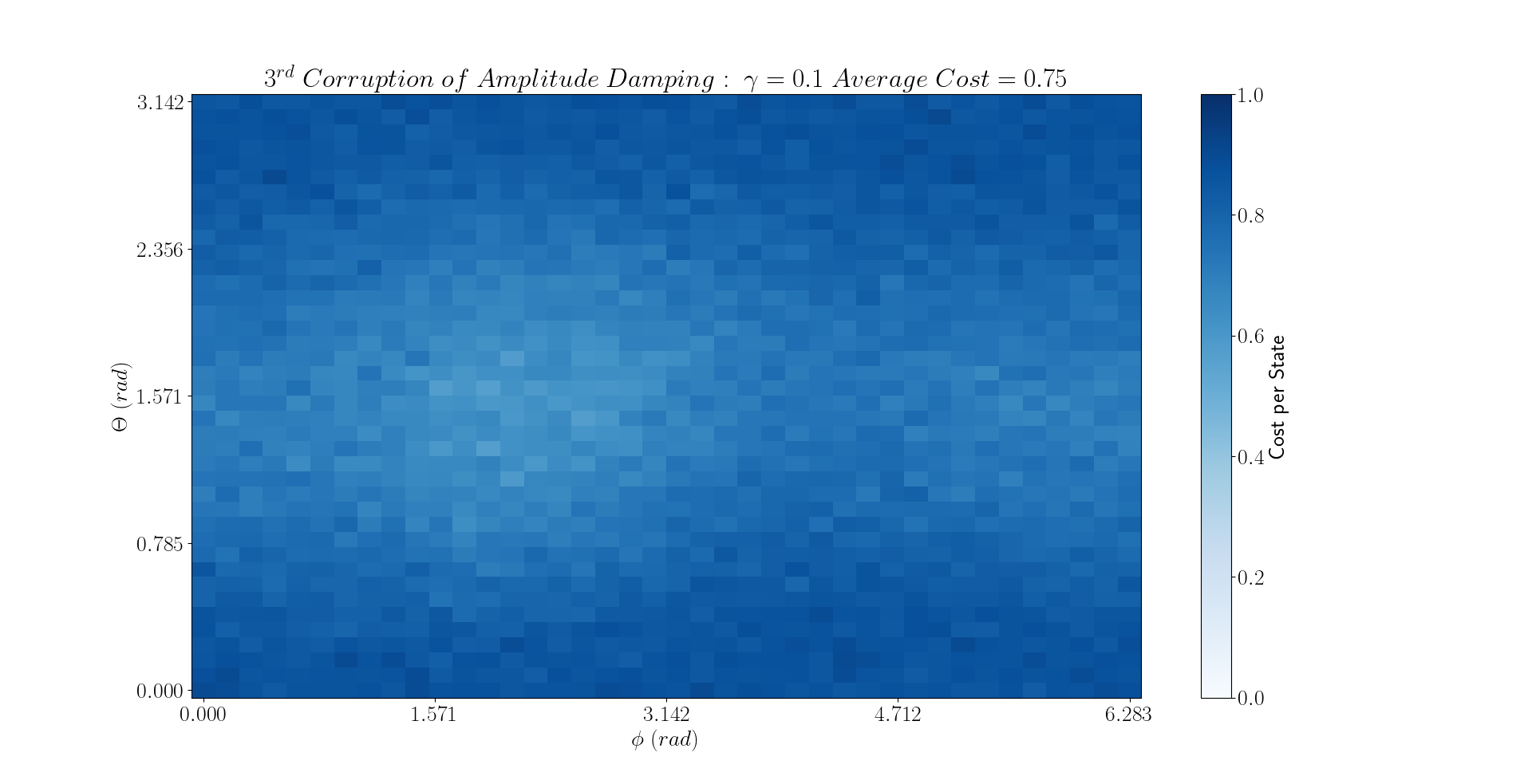}\\
    \includegraphics[width = 0.49\textwidth]{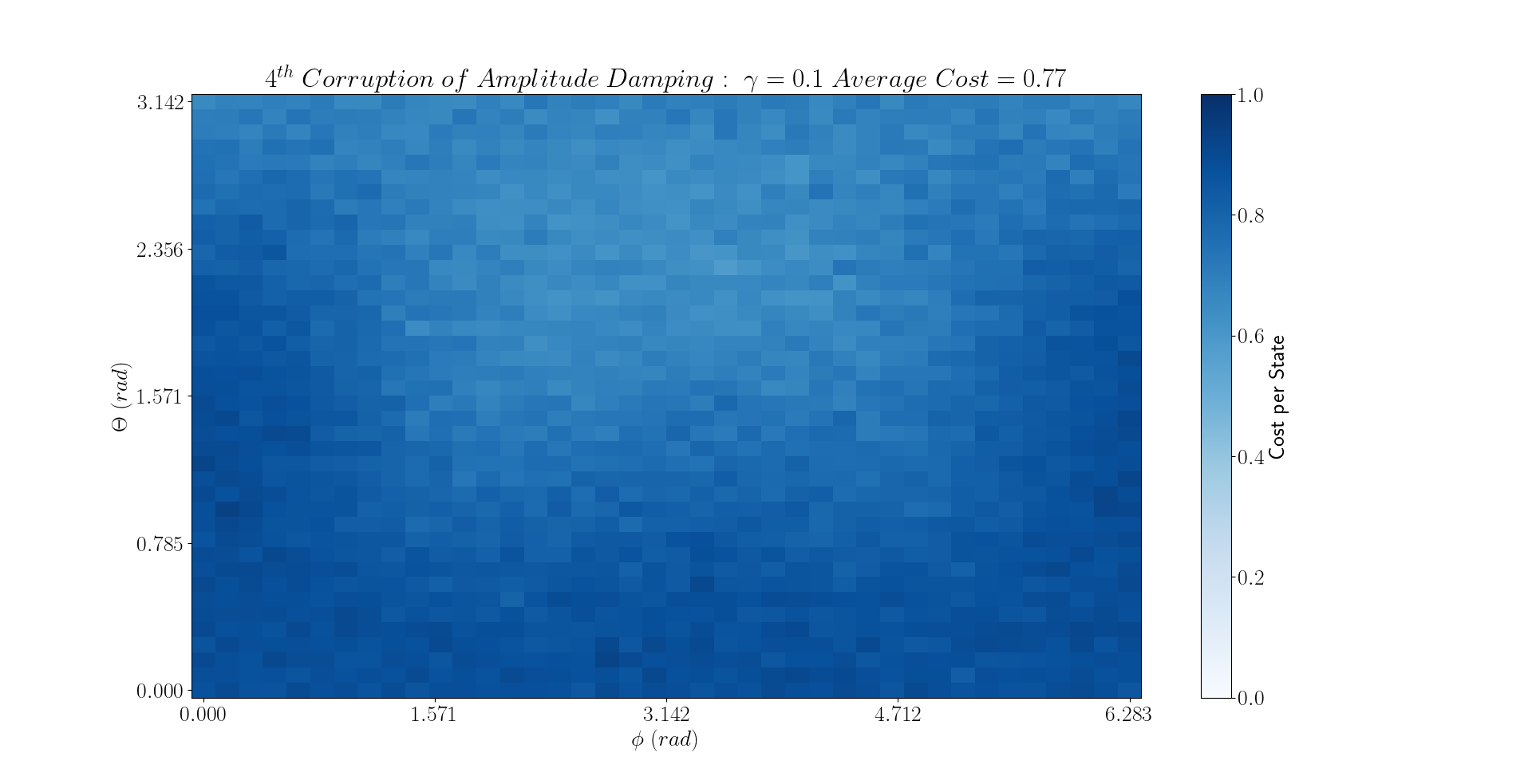}
    \includegraphics[width = 0.49\textwidth]{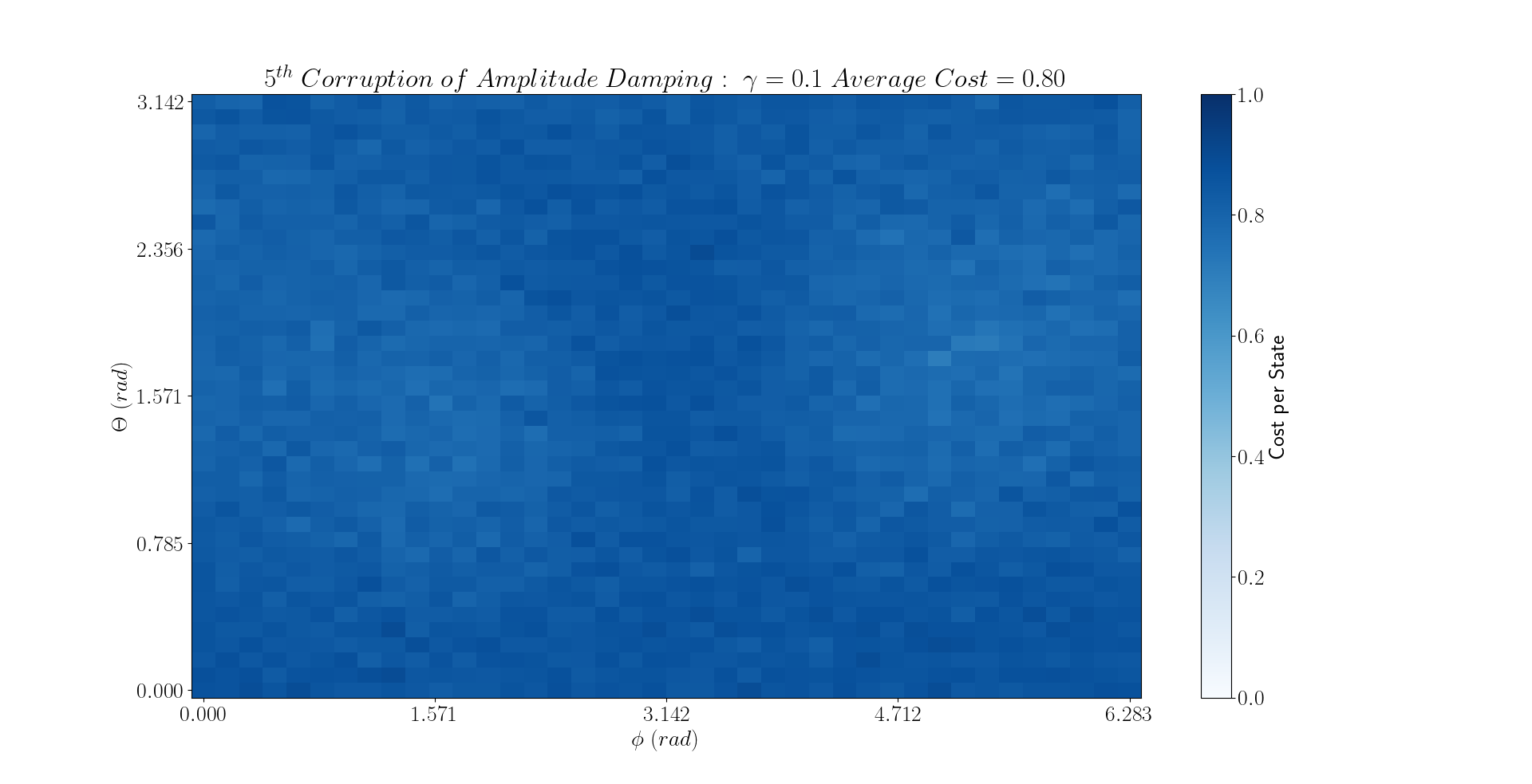}
    \caption{Results of the [4,1,4] QAE when error-correcting each one of the allowed corruptions of equation (\ref{eq6.19}) separately.}
    \label{fig6.14}
\end{figure}

Continuing with the testing of the model that produced the most satisfying results during its training, we examined its performance by error-correcting each one of the five possible corruptions that we have discussed in equation (\ref{eq6.19}) separately. To do so, we created 5 different validation sets that consist of 1600 different state pairs. Then, we exposed each validation set to a quantum operation that only allowed one of the possible 5 corruptions of Amplitude Damping to take place and we calculated the fidelity between the output of our model and the target state. Also, the average fidelity was calculated by using the equation (\ref{eq11}). We present our results in the form of colormaps, just like we did in the case of the Bit flip channel, in figure (\ref{fig6.14}).

The colormaps indicate that the performance of this model is not optimal. This QAE, with the average cost being $0.8$, cannot efficiently error-correct Amplitude Damping as we would expect the average cost to be no less than 0.95 for every possible corruption, as it was in the case of the Bit flip channel. Unfortunately, from this point on, no matter the combination of the hyperparameters that we used the average Fidelity either remained the same, or it started dropping. This phenomenon indicates that this QAE has reached a local maximum that prevents it from reaching the global maximum of the cost function that would allow our models to reach higher values of Fidelity. It is possible that models with different architectures or models that utilize larger training sets might perform better but our limited resources would not allow us to explore these options.

\section{Discovering Codewords with Quantum Neural Networks}

In previous sections, we used QNNs to denoise specific quantum noisy channels by training them on arbitrary qubit states. This was achieved by utilizing error-correcting codes that encode multiple physical qubits into a logical qubit, allowing the QAE to recognize and correct errors. An intriguing possibility is to let the QNN discover the error-correcting code on its own, which could potentially lead to the identification of previously unknown encodings for unconventional noisy channels.

\begin{figure}[h!]
    \begin{subfigure}{0.45\textwidth}
    \includegraphics[width = \textwidth]{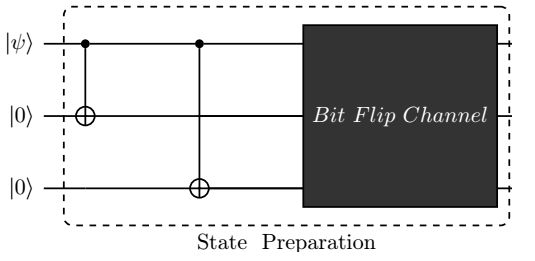}
    \caption{The state preparation is the part of the training quantum circuit which constructs the noisy quantum states $\tilde{\ket{\psi}}$ that we fitted into the QAE of section 5, in order to denoise them.}
    \label{state_prep}
    \end{subfigure}
    \begin{subfigure}{0.4\textwidth}
    \includegraphics[width = \textwidth]{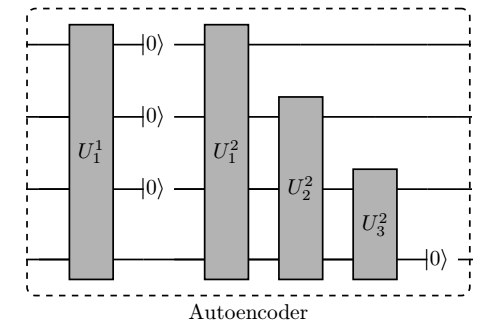}
    \caption{A [3,1,3] QAE used to denoise bit-flips}
    \end{subfigure}
    \begin{subfigure}{\textwidth}
        \includegraphics[width=\textwidth]{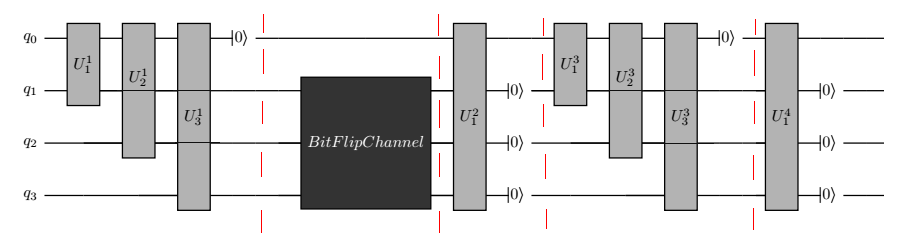}
        \caption{A [1,3,1,3,1] QNN with the Bit-Flip channel integrated into it. It combines the state preparation part of a QAE together with the QAE itself, into one. The last layer is added such that it performs the reverse transformation of layer 1.}
        \label{13131_QNN}
    \end{subfigure}
\end{figure}

To accomplish this, we propose a strategy that incorporates the noisy Quantum Channel into the neural network structure, rather than training the QAE on noisy quantum states as we have done before. In other words, we integrate the state preparation circuit (figure (\ref{state_prep})) that constructs and corrupts the logical qubit directly within the QNN. This enables the QNN to determine, during training, the optimal transformation for creating a logical qubit tailored to the specific quantum channel we've selected. With this approach, the QNN now consists of three main components:

\begin{itemize}
\item The first component, occurring at the initial layer, is the encoding process; the QNN needs to encrypt the input state into a multi-qubit logical state.
\item Next, the Noisy Quantum Channel corrupts the logical state that the QNN has generated.
\item Finally, the subsequent layers mimic the structure of a QAE, which we have shown to be effective in denoising corrupted logical states.
\end{itemize}

The first layer must have at least as many qubits as the theoretical minimum needed to construct the logical qubit for the specific quantum channel we've integrated. For instance, for the bit-flip channel, we need a minimum of 3 qubits. If the theoretical minimum is unknown for the quantum channel we've used, then the length of the first layer essentially becomes a hyperparameter. By attempting to reproduce its input, we believe the QNN will adapt to the integrated noise and eventually discover a quantum error-correcting code in the process.

As a proof of concept, we will begin by integrating the bit-flip channel inside the QNN, as we are familiar with it and know that the logical qubit requires only 3 (physical) qubits for implementation. The architecture of this QNN will be $[1,3,1,3,1]$ (see figure (\ref{13131_QNN})), with the bit-flip channel positioned between the first and second layers, immediately after the encryption of the input layer occurs. The next two layers form a QAE, as implemented in previous sections, and the output layer must perform the reverse transformation that constructs the logical qubit. Once training is complete, we can recover the 3-qubit codeword by examining the transformation of the first layer.

\begin{figure}[h!]
    \centering
    \includegraphics[width = 0.65 \textwidth]{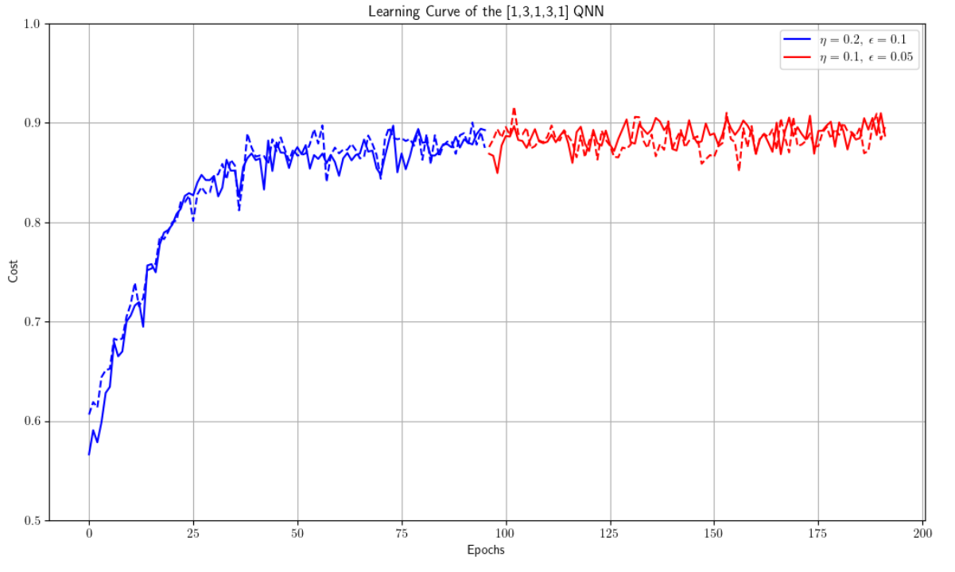}
    \caption{The dashed line indicates the cost function on the training set, while the continued one on the validation set. It is clear that this model reaches a barren plateau.}
    \label{arni}
\end{figure}

The learning curve for the first two sessions of the best-performing model can be observed in figure (\ref{arni}). The training set contained 100 input-target training pairs of the form $\qty{\ket{\phi},\ket{\phi}}$, where $\ket{\phi}$ is one of the following set of states: $\qty{\ket{0}, \ \ket{1}, \ \ket{+}, \ \ket{-}}$ (each state is equally likely to appear in the set). The validation set consisted of 400 states designed to cover the Bloch Sphere uniformly. The remaining hyperparameters included the use of stochastic gradient descent (SGD) as the optimizer and no batch size. The integrated bit-flip channel corrupts with a probability of $p=0.75$ (single-qubit errors), which was chosen to ensure that it outputs the following states with equal probability (otherwise the QNN would tend to develop a bias towards one of the 4 different cases)

\begin{equation}
    \begin{aligned}
        a\ket{000}+b\ket{111}, & \; \text{no bit-flip}.\\
	a\ket{100}+b\ket{011}, & \; \text{bit-flip to the 1st qubit}.\\
	  a\ket{010}+b\ket{101}, & \; \text{bit-flip to the 2nd qubit} \\
	a\ket{001}+b\ket{110}, & \; \text{bit-flip to the 3rd qubit}
\end{aligned}
 \label{bit_flip_ch}
\end{equation}

While the model initially seems to be learning, with the cost function eventually reaching a value of around 0.9, it plateaus without further improvement. This behavior is characteristic of all such models we tested, regardless of the chosen hyperparameters. Although a cost function value of 0.9 is not necessarily poor, the fact that most models rapidly converge to this point and stagnate prompts us to investigate a potential underlying reason.

\begin{figure}[h!]
    \centering
    \includegraphics[width=0.65\textwidth]{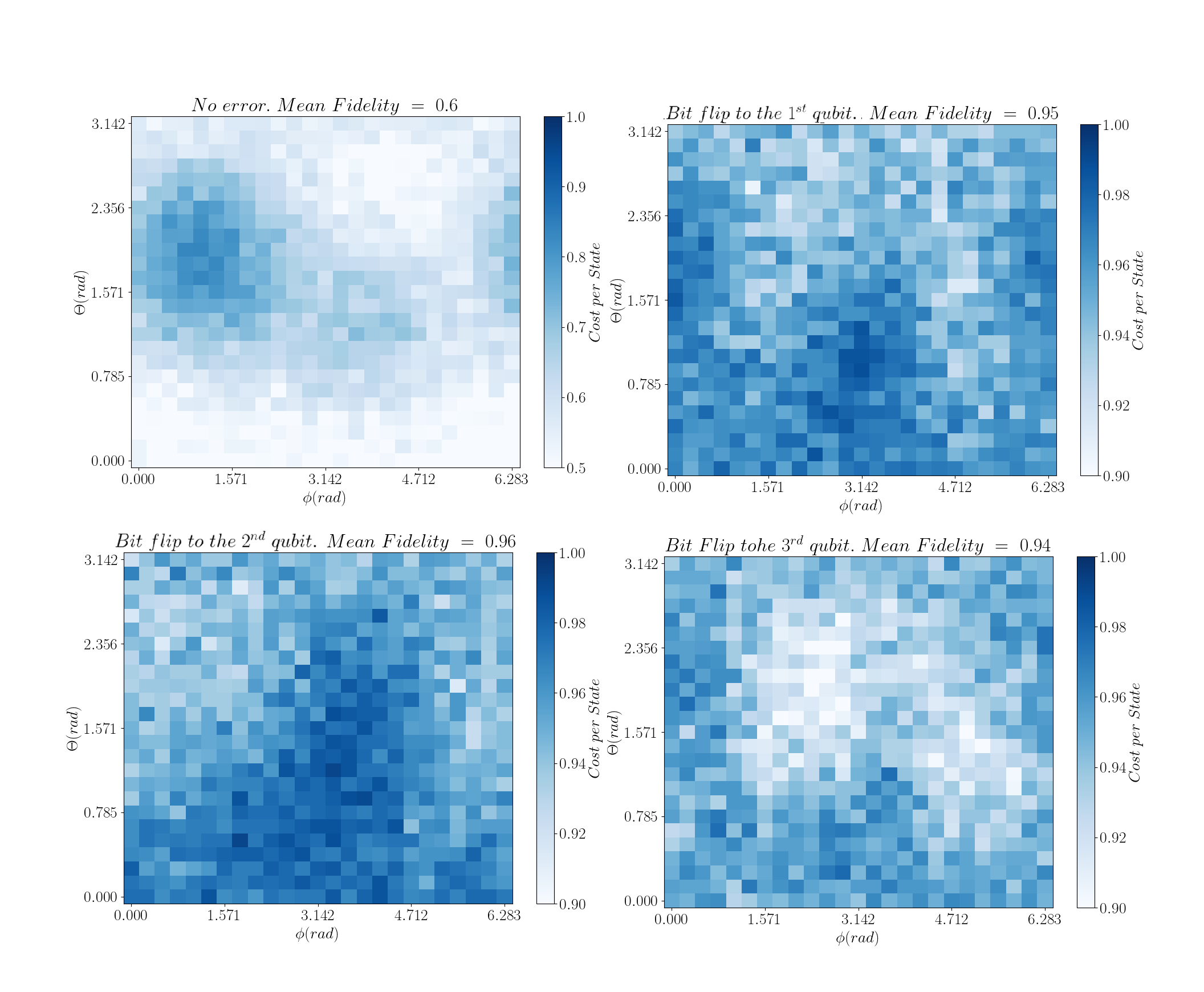}
    \caption{ The colormaps depict the fidelity, $F(\rho(\theta,\phi)^{out},\ket{\phi(\theta,\phi)}^{target})$, of output-target states of the [1,3,1,3,1] QNN on the Bloch sphere in the 4 cases that the bit-flip channel can affect the logical qubit. The QNN fails in cases where the bit flip channel doesn't corrupt the states, with its Mean Fidelity being 0.6.}
    \label{13131_bad}
\end{figure}

Further insight can be gained by examining the performance of the QNN for the four possible cases that the bit-flip channel can manifest. Figure \ref{13131_bad} illustrates the cost function on four validation sets, plotted as colormaps on the Bloch Sphere. In each validation set, the bit-flip channel either corrupts one of the three qubits or leaves them uncorrupted. A clear asymmetry is observed; the mean fidelity when bit-flip channel does not corrupt the state is way lower than the other cases, the QNN essentially fails to reconstruct its input despite explicitly setting the error probability, p, to balance out any potential bias of this kind. Moreover, this phenomenon is common among all other similar models we tested, with the only difference being that they exhibited this asymmetry with any of the four previously mentioned cases.

A possible explanation that can justify the above observations could be that the Cost function has 4 different local maximums, one for each case. Each time we set out to train a new model it randomly ends up in one such maximum thus preventing the cost function from reaching the global maximum.

\begin{figure}[h!]
    \begin{subfigure}{0.34\textwidth}
    \includegraphics[width = \textwidth]{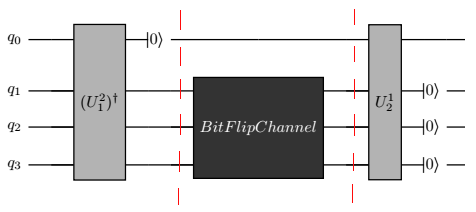}
    \caption{[1,3,1] QNN architecture.}
    \label{alkaida}
    \end{subfigure}
    \begin{subfigure}{0.65\textwidth}
    \includegraphics[width = \textwidth]{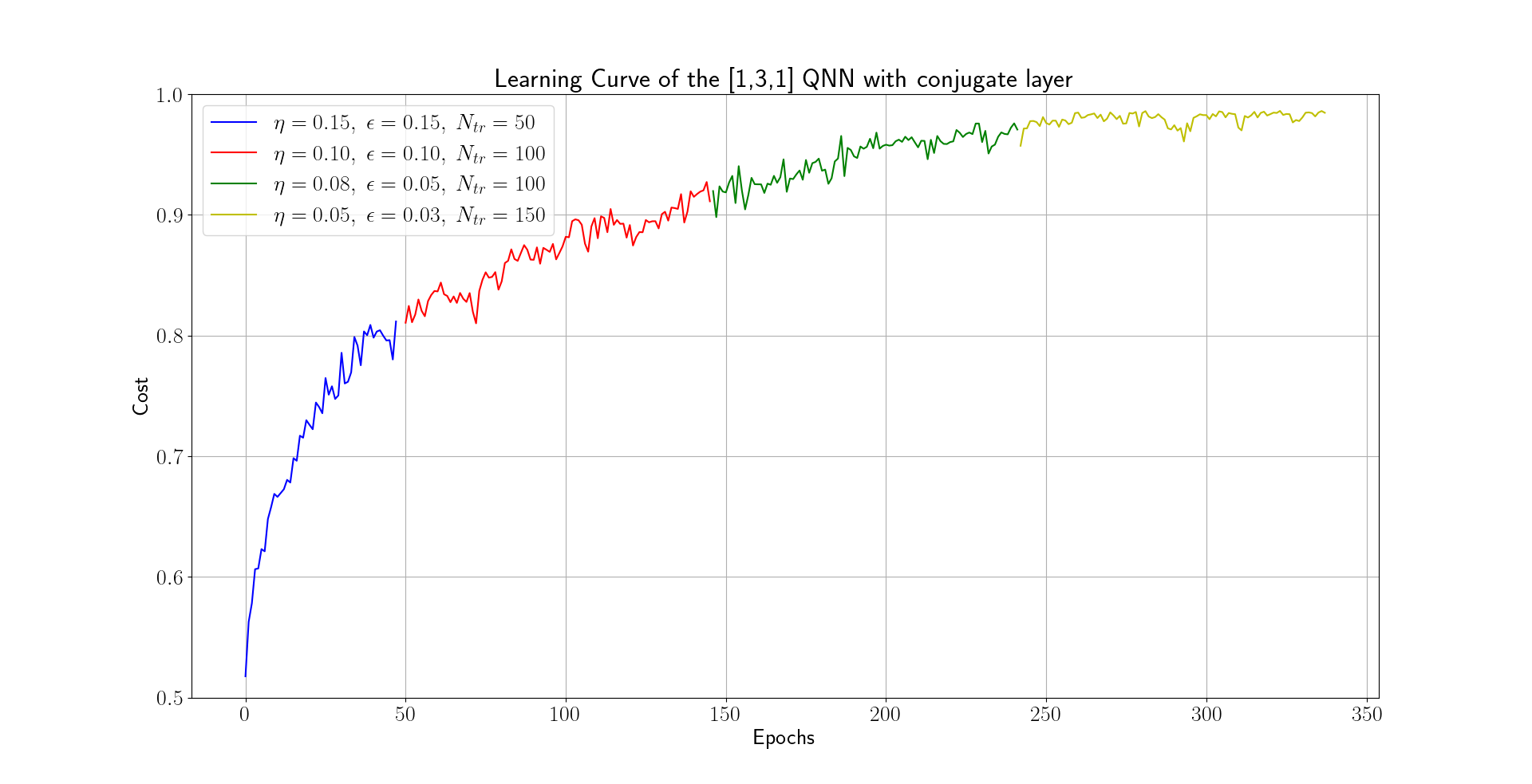}
    \caption{Learning curve of the [1,3,1] QNN with conjugate layer.}
    \label{131_learning_curve_conj}
    \end{subfigure}
    \begin{subfigure}{\textwidth}
        \centering
        \includegraphics[width=0.65\textwidth]{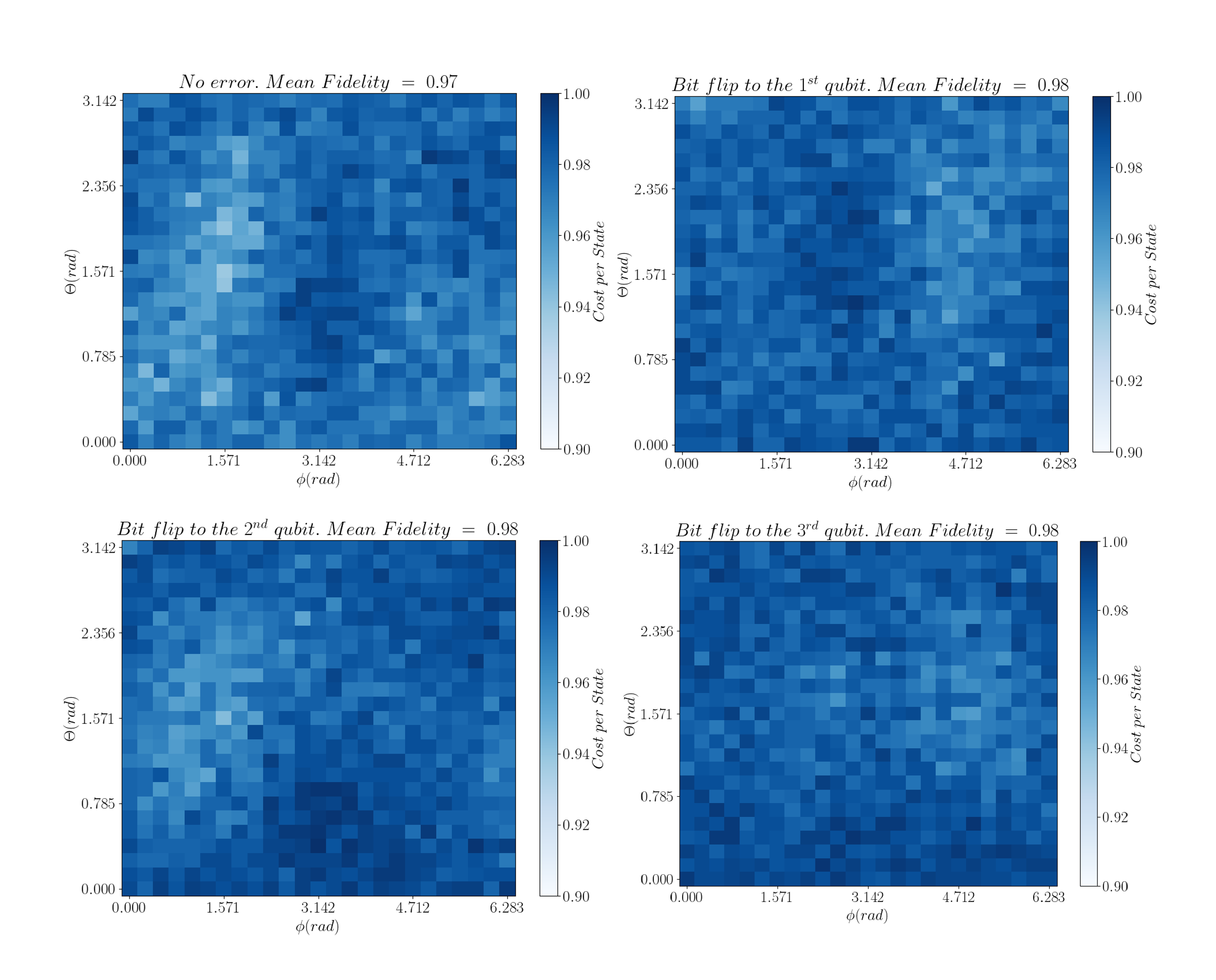}
        \caption{Same colormaps that figure (\ref{13131_bad}) depict, but with a $[1,3,1]$ QNN. This time the QNN retains high fidelity for all cases.  }
    \label{fig:my_label}
    \end{subfigure}
\end{figure}

By modifying the initially theorized structure of these QNNs and introducing conjugate layers into the architecture, we achieved more satisfactory results. This model features a simpler $[1,3,1]$ QNN structure, essentially a reverse autoencoder with the first layer acting as the conjugate of the second one, as illustrated in figure (\ref{alkaida}). In this design, we have effectively replaced two layers, whose original purpose was to replicate the denoising properties of a QAE, with a single conjugate layer. As a result, we have reduced the number of parameters in the network from 608 to 256, which not only made this model faster to train but also helped it surpass the 0.9 threshold value of the cost function that plagued previous models, as depicted in figure (\ref{131_learning_curve_conj}).

This effectively means that this model has successfully generated an error-correcting code specifically tailored for the bit-flip channel, maintaining high fidelity when denoised. The natural question that arises is: How does the encryption developed by the QNN compare to the conventional encryption used for the bit-flip channel? Since the encryption for the bit-flip code is not unique (alternative encryptions with similar properties can easily be found), we expect some differences. To investigate, we need to determine the transformation that the first layer performs on the basis $\qty{\ket{0},\ket{1}}$ and examine the resulting states. Doing so, we obtain:
$\qty{\ket{0},\ket{1}}$ and see how they are transformed. Doing this we get
\begin{equation}
    \ket{0}\rightarrow (0.92+i0.36 )\ket{001} + ... \ ,\quad \ket{1}\rightarrow (-0.81+i0.57 )\ket{011} + ... 
\end{equation}
where the coefficients following the ellipsis are very small and therefore not presented here. This result essentially confirms our intuition; the QNN has discovered a new encryption for the bit-flip channel, but it is still fundamentally based on the same principles we are familiar with for these codes. Specifically, the subspace $\qty{ \ket{001}, \ket{011} }$ is associated with cases of no error, while the other three subspaces correspond to each of the possible corruptions.

In conclusion, we demonstrated that a QNN can successfully generate an error-correcting code specifically tailored to address the bit-flip channel, without relying on a priori knowledge of a suitable encryption. By integrating the noisy quantum channel within the QNN and incorporating conjugate layers, we achieved a simpler and more efficient model, overcoming the limitations observed in previous attempts. Interestingly, the QNN-derived encryption, although distinct from conventional encryptions, still adheres to the same fundamental principles. This approach opens up new possibilities for discovering previously unknown encryptions for unconventional noisy channels, further expanding the capabilities and applications of quantum error-correcting codes in the field of quantum computing.

\section{Conclusions}

In this paper, we have explored the use of Quantum Autoencoders  and Quantum Neural Networks to design and implement quantum error-correcting codes for various quantum channels. Our investigation began with the bit-flip channel, where we successfully employed QAEs to error-correct corrupted states, further extending our approach to the more complex Amplitude Damping channel, which required the use of an approximative error-correcting code.

We then shifted our focus to a more ambitious goal: enabling QNNs to discover error-correcting codes without relying on a priori knowledge of suitable encryptions. By integrating the noisy quantum channel within the QNN and incorporating conjugate layers, we successfully demonstrated the ability of the QNN to generate an error-correcting code tailored to address the bit-flip channel. Interestingly, the derived encryption, although distinct from conventional encryptions, adhered to the same fundamental principles.

However, we observed that Dissipative Quantum Neural Networks can sometimes exhibit plateaus while training, which prevent them from reaching the global maximum of the cost function and consequently make it challenging to train them for specific purposes. This observation aligns with the findings in \cite{trainability}, where the authors characterize such QNNs as untrainable due to the barren plateaus they tend to fall into during training. While this may dampen the initial enthusiasm for the future of these networks, we believe that innovative solutions may exist to address this issue. In our case, we discovered that conjugate layers can offer some assistance in training these models while also accelerating the training process.

Our results indicate that QNNs have the potential to discover previously unknown encryptions for unconventional noisy channels, opening up new possibilities in the field of quantum computing. This work lays the foundation for future research into the development of novel quantum error-correcting codes and their applications in various quantum communication and computation scenarios. As quantum technology continues to advance, the ability to adapt and optimize error-correcting codes for different quantum channels will become increasingly crucial in ensuring reliable and efficient quantum systems. Creative approaches to overcome the challenges associated with DQNNs, such as barren plateaus, will be essential in unlocking their full potential.

\printbibliography

 \newpage
 \section*{Appendix A: Swap Test}

\begin{figure}[h!]
    \centering
    \includegraphics[width=0.75\textwidth]{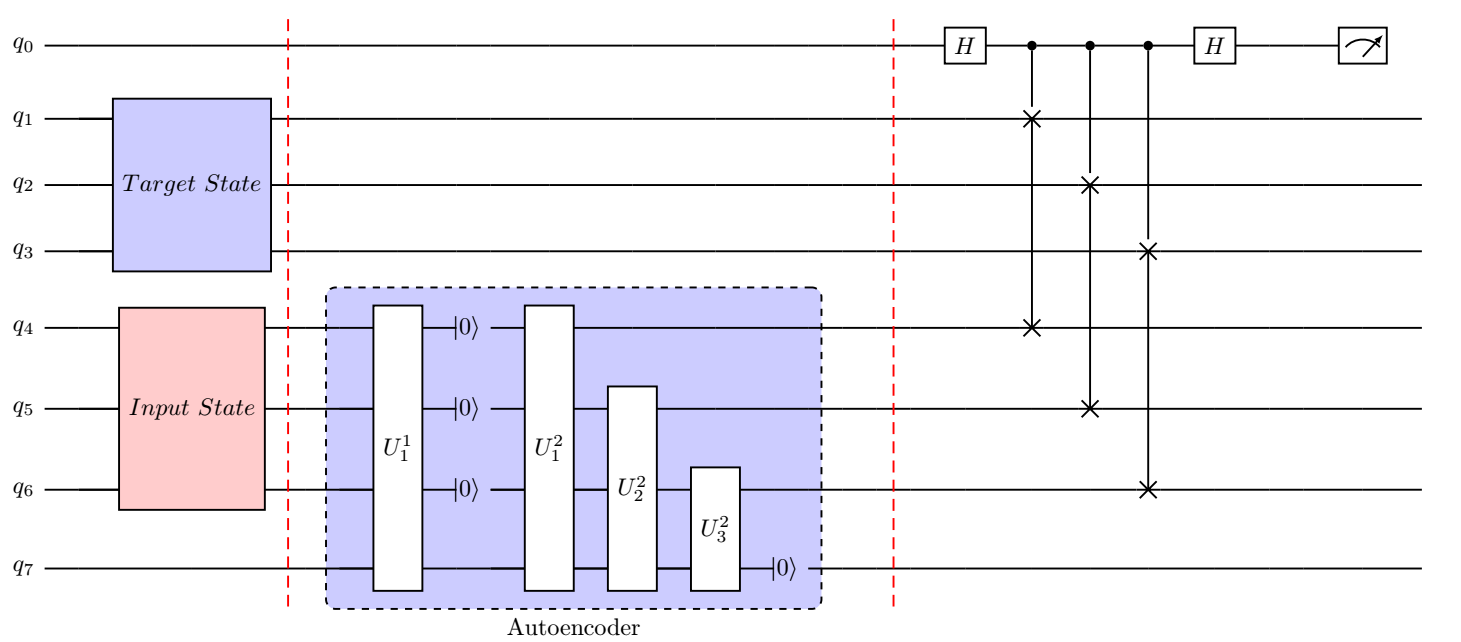}
    \caption{Quantum circuit that shows the 4 parts of the training process of a [3,1,3] QNN. A barrier separates each part. The first part prepares the pair $ \qty{\ket{\phi^{in}} ,\ket{\phi ^{targ}}}$. Then, the input state is transformed via the QNN. Afterward, a series of controlled swaps implement the Swap test. Lastly, the auxiliary qubit is measured and the outcome is saved into a classical register. }
    \label{qc_train_qnn}
\end{figure}

The Swap test essentially helps us perform a POVM at the output state of the QNN and the target state, in order to deduce their closeness. Let's briefly explain how this works. We will assume for simplicity that the output state of the network is pure $\ket{\phi^{out}}$. Starting from the second barrier of the circuit depicted in (\ref{qc_train_qnn}) let's follow a step-by-step of how the state of the first 7 qubits evolves.

Initially, the state is
\begin{equation}
    \ket{\Psi^{(0)}} = \ket{0}\ket{\phi^{targ}}\ket{\phi^{out}}
\end{equation}
The Hadamard gate on the first qubit, puts it in an equal superposition
\begin{align}
    \ket{\Psi^{(1)}} = \frac{(\ket{0}+\ket{1})}{\sqrt{2}}\ket{\phi^{targ}}\ket{\phi^{out}}= 
    \frac{1}{\sqrt{2}} \qty[ \ket{0}\ket{\phi^{targ}}\ket{\phi^{out}} + \ket{1}\ket{\phi^{targ}}\ket{\phi^{out}} ]
\end{align}
Next up, the series of controlled swaps essentially swap the state of the target qubits with that of the output qubits if the auxiliary is in the state $\ket{1}$
\begin{align}
     \ket{\Psi^{(2)}}=
    \frac{1}{\sqrt{2}} \qty[ \ket{0}\ket{\phi^{targ}}\ket{\phi^{out}} + \ket{1}\ket{\phi^{out}}\ket{\phi^{targ}} ]
\end{align}
Then, a Hadamard gate gets applied to the auxiliary qubit again 
\begin{align}
     \ket{\Psi^{(3)}}=\frac{1}{2} \ket{0} \qty[
   \ket{\phi^{targ}}\ket{\phi^{out}}+\ket{\phi^{out}}\ket{\phi^{targ}}] +\frac{1}{2} \ket{1} \qty[
   \ket{\phi^{targ}}\ket{\phi^{out}}-\ket{\phi^{out}}\ket{\phi^{targ}}]
\end{align}
Finally, the probability that the auxiliary qubit is 0 is given by 
\begin{align}
P_0 &= 
\frac{1}{2} \qty(
   \ket{\phi^{targ}}\ket{\phi^{out}}+\ket{\phi^{out}}\ket{\phi^{targ}}) \frac{1}{2}\qty(
   \ket{\phi^{targ}}\ket{\phi^{out}}+\ket{\phi^{out}}\ket{\phi^{targ}})^{\dagger} \\ 
   &= \frac{1}{2}+\frac{1}{2}|\langle \phi^{targ} \mid \phi^{out} \rangle|^2
\end{align}
If the output of the network is not pure but mixed, the steps to calculate this probability are very similar, and in the end, the Fidelity essentially replaces the inner product as the generalized version of it,
\begin{align}
P_0 = \frac{1}{2}+\frac{1}{2} F(\rho^{out},\ket{\phi^{targ}})
\end{align}
We can use this result to approximate the fidelity by re-arranging this relation and solve for it. What's left now is to estimate the probability $P_0$, which can be done by running the same circuit multiple times and then recording the times the auxiliary was measured in the state 0. We denote by $S$ the number of times we run each circuit, which has to be chosen to be big enough so that have a good enough approximation of $F$ but so big that we waste a lot of computational time unnecessarily. A good starting point is to choose $S=1000$ then once the QNN is at a plateau point or the cost function begins to oscillate between the same value, we may increase $S$ for more accuracy.

\end{document}